\documentclass[aps,prb,twocolumn,superscriptaddress,showpacs]{revtex4}
\usepackage{feyn}
\usepackage{graphicx}
\usepackage{latexsym}
\usepackage{amssymb}
\usepackage{amsmath}
\usepackage{amsfonts}
\usepackage{bm}
\usepackage{multirow}
\usepackage{color}
\usepackage{comment}
\newcommand{\ii}{\mathrm{i}}

\newcommand{\bx}{\mathbf{x}}

\newcommand{\cA}{\mathcal{A}}

\newcommand{\hn}{\hat{n}}
\newcommand{\htheta}{\hat{\theta}}

\newcommand{\tO}{\tilde{O}}
\newcommand{\tZ}{\tilde{Z}}

\newcommand{\SU}{\mathrm{SU}}
\newcommand{\U}{\mathrm{U}}

\newcommand{\beq}{\begin{equation}}
\newcommand{\eeq}{\end{equation}}
\newcommand{\beqn}{\begin{eqnarray}}
\newcommand{\eeqn}{\end{eqnarray}}

\DeclareMathAlphabet{\mathbbold}{U}{bbold}{m}{n}

\def\SU{{\rm SU}}

\def\U{{\rm U}}

\newcommand{\cC}{\mathcal{C}}

\newcommand{\lrar}{\leftrightarrow}

\begin{document}

\title{Universal Features of Higher-Form Symmetries at Phase Transitions}

%\author{UCSB CMT}

%\affiliation{Department of Physics, University of California,
%Santa Barbara, CA 93106, USA}

\author{Xiao-Chuan Wu}
\affiliation{Department of Physics, University of California,
Santa Barbara, CA 93106, USA}

\author{Chao-Ming Jian}
\affiliation{Department of Physics, Cornell University, Ithaca,
New York 14853, USA}

\author{Cenke Xu}
\affiliation{Department of Physics, University of California,
Santa Barbara, CA 93106, USA}

\begin{abstract}

We investigate the behavior of higher-form symmetries at various
quantum phase transitions. We consider discrete 1-form symmetries,
which can be either part of the generalized concept ``categorical
symmetry" (labelled as $\tZ_N^{(1)}$) introduced recently, or an
explicit $Z_N^{(1)}$ 1-form symmetry. We demonstrate that for many
quantum phase transitions involving a $Z_N^{(1)}$ or $\tZ_N^{(1)}$
symmetry, the following expectation value $ \langle \left( \log
O_\mathcal{C} \right)^2 \rangle$ takes the form $\langle \left(
\log O_\mathcal{C} \right)^2 \rangle \sim - \frac{A}{\epsilon} P+
b \log P $, where $O_\cC$ is an operator defined associated with
loop $\cC$ (or its interior $\cA$), which reduces to the Wilson
loop operator for cases with an explicit $Z_N^{(1)}$ 1-form
symmetry. $P$ is the perimeter of $\cC$, and the $b \log P$ term
arises from the sharp corners of the loop $\cC$, which is
consistent with recent numerics on a particular example. $b$ is a
universal microscopic-independent number, which in $(2+1)d$ is
related to the universal conductivity at the quantum phase
transition. $b$ can be computed exactly for certain transitions
using the dualities between $(2+1)d$ conformal field theories
developed in recent years. We also compute the ``strange
correlator" of $O_\cC$: $S_{\cC} = \langle 0 | O_\cC | 1 \rangle /
\langle 0 | 1 \rangle$ where $|0\rangle$ and $|1\rangle$ are
many-body states with different topological nature.

\end{abstract}

\maketitle

\section{Introduction}

The concept of symmetry is the most fundamental concept in
physics, and has profound implications and constraints on physical
phenomena. In recent years various generalizations of the concept
of symmetry have been explored. For example, ordinary symmetries
in a $d-$dimensional system are associated with the global
conservation of the symmetry charges, and the symmetry charges
localized within a $d-$dimensional subsystem of the space can only
change through the Noether current flowing across the surface of
the subsystem. In recent years the concept of 1-form symmetry
(more generally higher form symmetry) was proposed (see for
example
Ref.~\onlinecite{formsym0,formsym1,formsym2,formsym3,formsym4,formsym5,formsym6,formsym7,formsym8,Cordova2019}),
and the concept of 1-form symmetry is associated with conserved
``flux" through a $(d-1)-$dimensional subsystem; and the flux can
only change through the flowing of a 2-form symmetry current
across the edge of the $(d-1)-$dimensional subsystem. The concept
of 1-form symmetry was proven highly useful when analyzing gauge
fields. Using this new concept of symmetry and its 't Hooft
anomaly, it was proven that gauge fields with certain topological
term cannot be trivially gapped~\cite{formanomaly}, which is an
analogue of the Lieb-Shultz-Mattis theorem in condensed matter
systems~\cite{LSM,hastings}.

Lagrangians are often used to describe a physical system, and the
form of the Lagrangian depends on one's choice of ``local degrees
of freedom" of the system, and other degrees of freedom may become
nonlocal topological defects in the Lagrangian. When we select
another set of local degrees of freedom of the same system to
construct the Lagrangian, it will take a new form, and the new
form of Lagrangian is related to the original Lagrangian through a
``duality transformation". It was realized in recent years that,
in some examples, duality transformation of the Lagrangian, along
with the explicit symmetry of the Lagrangian, could be embedded
into a larger symmetry group~\cite{xudual1,xudual2}, which may
only emerge in the infrared limit, and is not explicit unless one
takes into account of all the dual forms of the Lagrangian.

Most recently a new generalization of symmetry was developed which
also involves the dual description of a system. It is well-known
that certain models of theoretical physics have a dual
description, and the dual model has symmetries that are inexplicit
in the original model. A concept called ``categorical symmetry"
was developed which unifies the explicit symmetry of a system and
the inexplicit symmetry of its dual model, and treat them on an
equal footing~\cite{ji2019}. To diagnose the behavior of the
categorical symmetries, and most importantly to diagnose the
explicit symmetry and the inexplicit dual symmetry on an equal
footing, a concept of ``order diagnosis operator" (ODO) was
introduced, whose expectation value reduces to the correlation
function between order parameters for an explicit 0-form symmetry,
and reduces to a Wilson loop for an explicit 1-form
symmetry~\cite{wucat}. The ODO was also referred to as the patch
operator in Ref.~\onlinecite{ji2019}. For example, the ODO for the
$Z_2$ symmetry of the $2d$ quantum Ising model is $O_{ij} =
\sigma^z_i \sigma^z_j$, while the ODO for the dual $\tZ_2^{(1)}$
1-form symmetry is $\tO_{\cC} = \prod_{j \in \cA, \partial \cA =
\cC} \sigma^x_j$, where $\sigma^z$ transforms under the explicit
$Z_2$ symmetry. $\tO_{\cC}$ creates a domain wall of $\sigma^z$
along a closed loop $\cC$ by flipping the sign of $\sigma^z$ on a
patch $\cA$, which is the interior of $\cC$. ODOs for systems with
special symmetries such as subsystem symmetries may have special
forms and behaviors, and examples with these special symmetries
were discussed in Ref.~\onlinecite{wucat}.

The expectation value of $O_{ij}$ and $\tO_{\cC}$ in the $2d$
quantum Ising system characterizes different phases of the system.
In the two gapped phases, i.e. the ordered and disordered phase of
$\sigma^z$, the behavior of $\langle O_{ij} \rangle$ and $\langle
\tO_{\cC} \rangle$ are relatively easy to evaluate, since they can
be computed through perturbation~\cite{susskind}, which is
protected by the gap of the phases. In the ordered phase of
$\sigma^z$, $\langle O_{ij} \rangle$ saturates to a constant when
$|i - j| \rightarrow \infty$, and $\langle \tO_{\cC} \rangle$
decays with an area law; in the disordered phase of $\sigma^z$,
$\langle O_{ij} \rangle$ decays exponentially with $|i - j|$,
while $\langle \tO_{\cC} \rangle$ decays with a perimeter law. But
at the critical point of the system, i.e. the $(2+1)d$ quantum
Ising phase transition, the behavior of the ODO $\tO_{\cC}$ is
more difficult to evaluate. Ref.~\onlinecite{chengcat} evaluated
$\langle \tO_{\cC} \rangle$ numerically, and the result indicates
that in addition to a leading term linear with the perimeter of
$\cC$, a subleading term which is logarithmic of the perimeter
arises for a {\it rectangular} shaped loop $\cC$. The logarithmic
subleading contribution may be a universal feature of ODO at a
critical point, and the $Z_2$ ODO can be mapped to the 2nd Renyi
entanglement entropy of a {\it free} boson/fermion
system~\cite{chengcat}. Tt is known that there is a corner induced
logarithmic contribution for the Renyi entropy in a general
conformal field theory~\cite{eeloga,eelogb,eelog1,eelog2}.
However, for {\it interacting} systems the exact relation between
entanglement entropy and ODO is not clear yet.

In this work we demonstrate that, for a $2d$ quantum system with
either an explicit 1-form symmetry $Z_N^{(1)}$, or an inexplicit
symmetry $\tZ_N^{(1)}$ (which is dual to a 0-form ordinary $Z_N$
symmetry), the following quantity $\langle (\log O_{\cC})^2
\rangle$ or $\langle (\log \tO_{\cC})^2 \rangle$ take a universal
form $- \frac{A}{\epsilon} P + b \log P$ at many quantum critical
points. Here $P$ is the perimeter of the loop $\cC$. $b$ is a
universal number which arises from a sharp angle of the loop
$\cC$; $b$ is proportional to the universal conductivity of the
$2d$ quantum critical point, and it is a universal function of the
angle $\theta$. We demonstrate this result for various examples of
quantum critical points. We also comment on the connection between
ODO and entanglement entropy in the end of the manuscript.

We also compute a quantity called the ``strange correlator" of the
1-form ODO $O_{\cC}$. The strange correlator was introduced as a
tool to diagnose the symmetry protected topological (SPT) states
based on the bulk wave function instead of the edge
states~\cite{xusc}, and it was shown to be effective in many
examples~\cite{sengupta1,sengupta2,sengupta3,zohar1,zohar2,zohar3,mengstrange,beach,frank}.
In the current work we study the strange correlator for one
example of 1-form SPT state, but we expect similar studies are
worth pursuing for more general cases.

\section{Systems with dual $\tZ_N^{(1)}$ 1-form symmetry}

\subsection{Example 1: $Z_N$ order-disorder transition}

We first consider cases when the system has an explicit $Z_N$
(0-form) symmetry, and it has an inexplicit dual $\tZ_N^{(1)}$
1-form symmetry. The simplest example of quantum phase transition,
is the order-disorder transition of the $Z_N$ symmetry. The
lattice model with $Z_N$ symmetry, can be embedded into an
ordinary $\U(1)$ rotor model: \beqn H = \sum_{<i,j>} - t
\cos(\htheta_i - \htheta_j) + V(\hn_i) - 2u \cos(N\htheta_i),
\label{H} \eeqn where $[\hn_i, \ \htheta_j] = \ii \delta_{ij}$,
and $\htheta_j$ prefers to take values $\htheta_j = 2\pi k / N$
with $k = 0, \cdots N-1$ due to the $u$-term. The potential
$V(\hn)$ has a minimum at $\hn = 0$. The order-disorder transition
of the $Z_N$ symmetry is described by the Landau-Ginzburg action
\beqn \mathcal{S} &=& \int d^2x d\tau \ |\partial \Phi|^2 +
r|\Phi|^2 + g |\Phi|^4 + u (\Phi^N + h.c.) \lrar \cr\cr
\mathcal{S}_d &=& \int d^2x d\tau \ |(\partial - \ii a) \phi|^2 +
\tilde{r} |\phi|^2 + \tilde{g} |\phi|^4 \cr\cr &+& u (\mathrm{M}^N
+ h.c.). \label{dual1} \eeqn $\Phi$ is the complex order
parameter. The second line of the equation is the well-known
boson-vortex dual description of the phase
transition~\cite{peskindual,halperindual,leedual}, and $r \sim -
\tilde{r}$ is the tuning parameter of the transition: $r
> 0$ ($ r < 0$) corresponds to the gapped (condensed) phase of
$\Phi$ and condensed (gapped) phase of $\phi$. The $\Phi^N$ term
is the $Z_N$ anisotropy on $\Phi$ which breaks the $\U(1)$
symmetry of $\Phi$ to $Z_N$. The $\Phi^N$ is dual to the $N-$fold
monopole operator ($\mathrm{M}^N$) in the dual theory. It is known
that when $N \geq 4$, the $u$ term ($Z_N$ anisotropy) is an
irrelevant perturbation at the $(2+1)d$ XY transition, and there
will be an emergent $\U(1)$ symmetry at the quantum phase
transition.

As was discussed before, a system with $Z_N$ symmetry has an
inexplicit dual $\tZ_N$ 1-form symmetry, the $Z_N$ and
$\tZ_N^{(1)}$ symmetry together constitute the ``categorical
symmetry" of the system~\cite{ji2019}. In order to describe the
behavior of the $\tZ_N^{(1)}$ symmetry, Ref.~\onlinecite{wucat}
introduced the ``order diagnosis operator" $\tO_{\cC}$.
Represented in terms of lattice operators, the ODO for the dual
$Z_N^{(1)}$ symmetry reads \beqn \tO_{\cC} = \exp\left( \ii
\frac{2\pi}{N} \sum_{j \in \cA} \hn_j \right), \label{znodo} \eeqn
where $\partial \cA = \cC$ is a patch of the $2d$ lattice enclosed
by contractible loop $\cC$, and the ODO was also called patch
operator in Ref.~\onlinecite{ji2019}. $\tO_\cC$ creates a $Z_N$
domain wall. In the ordered and disordered phase of the $Z_N$
symmetry, the expectation value of $\tO_\cC$ decays with an area
law and perimeter law respectively.

At the order-disorder phase transition, to extract the universal
feature of the ODO $\tO_\cC$, we evaluate $ \langle ( \log
\tO_\mathcal{C})^2 \rangle$~\footnote{$\log$ is a multivalued
function. Since $\tO_\cC = \prod_j \tO_{j \in \cA, \partial \cA =
\cC}$, where $\tO_{j} = e^{\ii 2\pi \hn_j/N}$, we define $\log
\tO_\mathcal{C} = \sum_{j \in \cA} \log \tO_{j}$, and demand
Arg$[\tO_{j}] = \log \tO_j \in (-\pi, \pi] \sim 2\pi \hn_j/N$, the
$V(\hn_i)$ term in the Hamiltonian Eq.~\ref{H} restricts $\hn_j$
to largely fluctuate around its minimum $\hat{n}_j \sim 0$. },
which in the dual theory reduces to \beqn \langle (\log
\tO_\mathcal{C})^2 \rangle = - \frac{1}{N^2} \int_\cC dl^\mu
\int_{\cC'} dl^{\prime \nu} \langle a_\mu(\bx) a_\nu(\bx')
\rangle. \label{logznodo} \eeqn The relation between $a_\mu$ and
the original Landau-Ginzburg theory is $J = \frac{\ii}{2\pi} \ast
da$, where $J$ is the current of the emergent $\U(1)$ symmetry at
the $Z_N$ order-disorder transition. The correlation of $a_\mu$ is
dictated by the correlation of $J$ whose scaling dimension {\it
does not} renormalize at a general conformal field theory. The
correlation between currents $J$ is proportional to the universal
conductivity at a $(2+1)d$ conformal field theory: \beqn \langle
J_{\mu}(0)J_{\nu}(\mathbf{x})\rangle=
\sigma\frac{I_{\mu\nu}(\mathbf{x})}{\left|\mathbf{x}\right|^{4}},
\eeqn where the matrix $I_{\mu\nu}(\mathbf{x})$ is given by $
I_{\mu\nu}(\mathbf{x}) = \delta_{\mu\nu} -
2x_{\mu}x_{\nu}/\left|\mathbf{x}\right|^{2}$, and $\sigma$ is
$C_J$ in (for example) Ref.~\onlinecite{Giombi_2016}. The
universal conductivity at a $(2+1)d$ XY transition was predicted
in Ref.~\onlinecite{UC1}, and it can be computed using various
theoretical and numerical methods, and also measured
experimentally (see for example
Ref~\onlinecite{UCt1,UCt2,UCt3,UCt4,UCe1,UCe2,UCe3}, the universal
conductivity in some of the references was computed/measured with
strong disorder).

It is straightforward to verify that the gauge field propagator
can be written as \beqn \left\langle
a_{\mu}(0)a_{\nu}(\mathbf{x})\right\rangle
=\sigma\pi^{2}\frac{\delta_{\mu\nu}-\zeta
I_{\mu\nu}(\mathbf{x})}{\left|\mathbf{x}\right|^{2}}, \label{gauge
propagator} \eeqn The parameter $\zeta$ is introduced by a
nonlocal gauge fixing term \beqn
\frac{1}{8\pi^{6}\sigma}\frac{1}{1-\zeta}\int
d^{3}\mathbf{x}d^{3}\mathbf{y}\frac{\partial_{\mu}a^{\mu}(\mathbf{x})
\partial_{\nu}a^{\nu}(\mathbf{y})}{\left|\mathbf{x}-\mathbf{y}\right|^{2}},
\eeqn which contributes to a total derivative
$I_{\mu\nu}(\mathbf{x})/\left|\mathbf{x}\right|^{2}=
\frac{1}{2}\partial_{\mu}\partial_{\nu}\log\left|\mathbf{x}\right|^{2}$
in the gauge field propagator.

In the explicit calculation of Eq.~\ref{logznodo}, one should be
very careful about how to set the UV cut-off. A hard cut-off on
the integration interval $|\bx - \bx'|$ along $\mathcal{C}$ will
spoil the gauge invariance. To guarantee that $\cC$ and $\cC'$ are
both complete loops in the integral (hence gauge invariance is
preserved), a good method is to set a small distance between $\cC$
and $\cC'$ along the temporal direction by distance $\tau =
\epsilon>0$, and this small splitting serves as a small real-space
UV cut-off. The integral is then performed along the closed loop
$\mathcal{C}$ (and its duplicate $\cC'$) in the $x$-$y$ plane. For
a smooth loop $\cC$ with perimeter $P$, the evaluation of $
\langle \left( \log O_\mathcal{C} \right)^2 \rangle$ simply yields
a perimeter law, $i.e.$ proportional to $P$ with a UV-dependent
coefficient. For example, when $\cC$ is a circle with radius $R$,
the integral in Eq.~\ref{logznodo} gives \beqn
-\langle(\log\tilde{O}_{\mathcal{C}})^{2}\rangle=
\frac{\sigma\pi^{2}}{N^{2}}\left(\frac{2\pi^{2}R}{\epsilon}-2\pi^{2}
+\frac{3\pi^{2}\epsilon}{4R}\right)+\mathcal{O}(\epsilon^{2}).
\eeqn There are two observations. First, the final result is
independent of the gauge choice $\zeta$. Second, the large-$R$
scaling is only given by a linear term which depends on the UV
cut-off.

However, if the loop $\cC$ has sharp corners, the situation is
very different, and some universal feature that does not depend on
the UV cut-off emerges. Let us first consider $\cC$ being a
spatial square with four corners
$\left(0,0\right),\left(L,0\right),\left(L,L\right),\left(0,L\right)$.
There are three types of integrals that are involved. The linear
contribution is from the correlation along the same edge of $\cC$
\begin{flalign}
&\int_{0}^{L}dx\int_{0}^{L}dx^{\prime}\frac{(1+\zeta)(x-x^{\prime})^{2}
+(1-\zeta)\epsilon^{2}}{((x-x^{\prime})^{2}+\epsilon^{2})^{2}}
\notag
\\=\;&\frac{\pi L}{\epsilon}-2(1+\zeta)\log(L/\epsilon)+\mathcal{O}(1).
\label{linear and log from same line}
\end{flalign}
It is important to notice that there is a $\log(L/\epsilon)$ term,
which also shows up in the integral for two neighboring edges that
are perpendicular to each other
\begin{flalign}
\int_{0}^{L}dx\int_{0}^{L}dy^{\prime}\frac{2\zeta
xy^{\prime}}{(x^{2}+y^{\prime2}+\epsilon^{2})^{2}}=\zeta\log(L/\epsilon).
\label{log from right angle}
\end{flalign}
The integral from two parallel edges is a finite number which does
not grow with $L$
\begin{flalign}
\int_{0}^{L}dx\int_{0}^{L}dx^{\prime}
\frac{(\zeta+1)(x-x^{\prime})^{2}+(1-\zeta)(L^{2}+\epsilon^{2})}
{-(L^{2}+(x-x^{\prime})^{2}+\epsilon^{2})^{2}}=\mathcal{O}(1)
\end{flalign}
Combining all contributions together, we find the gauge invariant
result
\begin{flalign}
-\langle(\log\tilde{O}_{\mathcal{C}})^{2}\rangle=
\frac{\sigma\pi^{2}}
{N^{2}}\left(\frac{\pi4L}{\epsilon}-8\log(L/\epsilon)\right)+\mathcal{O}(1).
\end{flalign}
The $\zeta$-independence of the $\mathcal{O}(1)$ term has also
been verified. This result is similar to the evaluation of a
square Wilson loop for free QED in $(3+1)$ dimensions. In both the
two cases above, we find that the linear term in
$-\langle(\log\tilde{O}_{\mathcal{C}})^{2}\rangle$ is
$\frac{\sigma\pi^{2}}{N^{2}}\frac{\pi P}{\epsilon}$ where $P=2\pi
R$ for the circle and $P=4L$ for the square.

\begin{figure}
\includegraphics[width=0.3\textwidth]{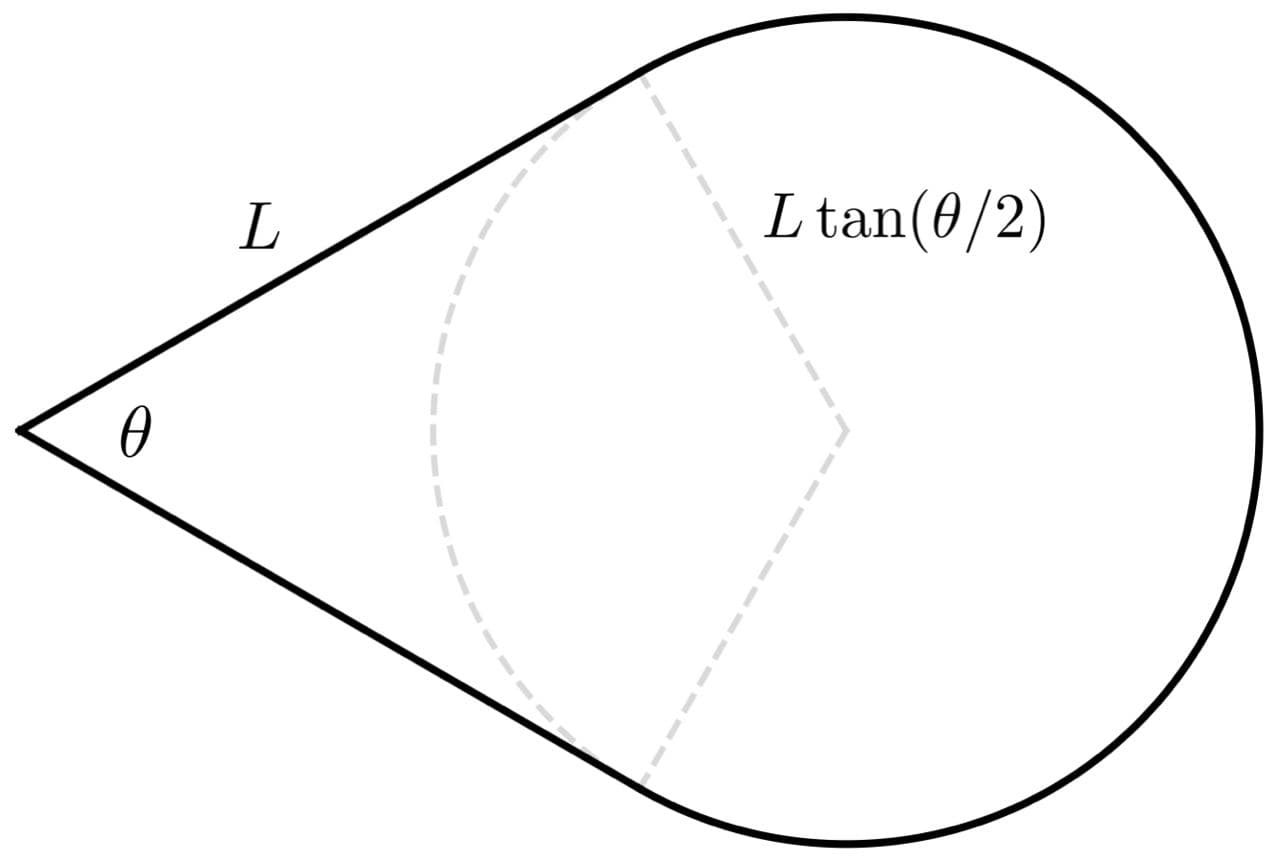}
\caption{The shape of $\cC$ with only one angle $0<\theta<\pi$. As
a concrete example, we consider a circle with two tangent lines
that intersect at a point. Each tangent line has the length $L$,
the radius of the circle is therefore $L\tan(\theta/2)$ and the
perimeter of $\cC$ is given by
$P=(2+(\pi+\theta)\tan(\theta/2))L$.} \label{AngleShape}
\end{figure}

\begin{figure}
\includegraphics[width=0.48\textwidth]{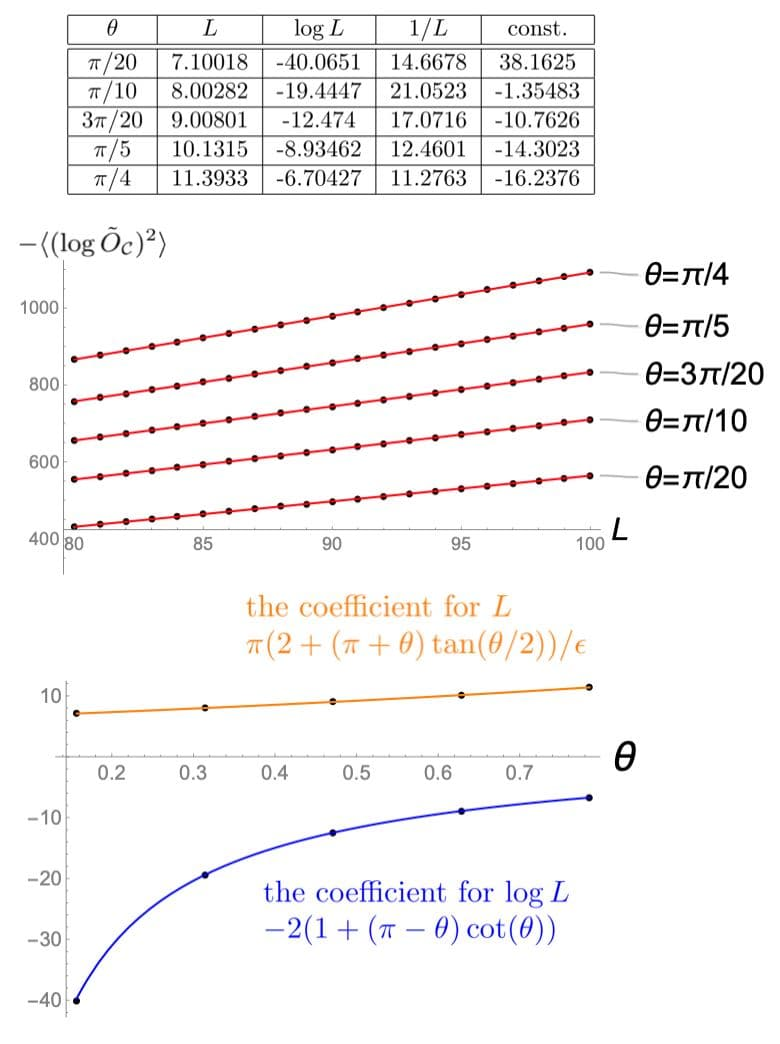}
\caption{The numerical results of
$-\langle(\log\tilde{O}_{\mathcal{C}})^{2}\rangle$ (in the unit of
$\sigma\pi^2/N^2$) for the shape in FIG.~\ref{AngleShape} with
different angles. The UV cut-off is set to be $\epsilon=1$. The
large-$L$ scaling is fitted by the function
$-\langle(\log\tilde{O}_{\mathcal{C}})^{2}\rangle=aL/\epsilon+b\log
L+c/L+d$, and the fitting parameters $a,b$ agree with the
analytical expressions Eq.~\ref{ODO with angle} and Eq.~\ref{angle
function}.} \label{AngleNumerical}
\end{figure}

Let us now generalize Eq.~\ref{log from right angle} to the case
of two straight lines with an arbitrary angle $\theta$ with $0 <
\theta < \pi$. For convenience, we choose the gauge $\zeta=0$ in
the following calculations. We could parametrize the two straight
lines by $t(\cos(\theta/2),-\sin(\theta/2))$ and
$s(\cos(\theta/2),\sin(\theta/2))$ where $0<s,t<L$. To extract the
angle-dependence of the logarithmic divergence, we use the trick
in Ref.~\onlinecite{angle1,angle2}
\begin{flalign*}
&\int_{0}^{L}ds\int_{0}^{L}dt\frac{-\cos\theta}{s^{2}+t^{2}-2st\cos\theta
+\epsilon^{2}}=\int_{0}^{L}d\ell\int_{0}^{1}d\lambda
\\&\left[\frac{\ell}{\ell^{2}+\epsilon^{2}}
\frac{-\cos\theta}{\lambda^{2}+(1-\lambda)^{2}-2\lambda(1-\lambda)\cos\theta}
+\mathcal{O}(\epsilon^{2}/\ell^{3})\right],
\end{flalign*}
where we have changed the integration variables to
$s=\ell\lambda,t=\ell(1-\lambda)$, and the
$\mathcal{O}(\epsilon^{2}/\ell^{3})$ part does not contribute to
any logarithmic divergence. The $\lambda$-integral can be
evaluated exactly, which gives $-(\pi-\theta)\cot\theta$. The
$\log(L/\epsilon)$ divergence then arises from the
$\ell$-integral.
%Although Eq.~\ref{log from right angle} vanishes
%with $\zeta=0$, there is a nonzero contribution for
%$\theta\neq\pi/2$. As shown in Eq.~\ref{linear and log from same
%line},
There is another logarithmic contribution from correlation within
the same line. Combining all the contributions together,
eventually we obtain
\begin{flalign}
-\langle(\log\tilde{O}_{\mathcal{C}})^{2}\rangle&=
\frac{\sigma\pi^{2}}{N^{2}}\left(\frac{\pi
P}{\epsilon}-f(\theta)\log P\right)+\mathcal{O}(1) \label{ODO with
angle} \\ f(\theta)&=2(1+(\pi-\theta)\cot(\theta)) \label{angle
function}
\end{flalign}
for any shape of $\cC$ with a single corner, where $P$ is the
perimeter of $\cC$. We observe that the universal logarithmic term
vanishes when $\theta=\pi$, and only the linear term remains, as
expected. To double check the analytical expression Eq.~\ref{ODO
with angle}, we consider the shape of $\cC$ as shown in
FIG.~\ref{AngleShape}, and the numerical evaluation for
$-\langle(\log\tilde{O}_{\mathcal{C}})^{2}\rangle$ for different
angles are shown in FIG.~\ref{AngleNumerical}. For fixed values of
$L,\epsilon$, the angle dependence for both the linear and the
logarithmic terms agree with Eq.~\ref{ODO with angle} and
Eq.~\ref{angle function}.

We computed $ - \langle (\log \tO_\cC)^2 \rangle$, which is the
second order expansion of $2 \langle \tO_\cC \rangle$. We have not
proven whether higher order expansion in $\langle \tO_\cC \rangle$
leads to different corner contribution from $\langle (\log
\tO_\cC)^2 \rangle$ or not. We would also like to mention that the
entanglement entropy of a patch $\cA$ with corners in a $(2+1)d$
CFT is related to another universal quantity $C_T$ from the
correlation of the stress-energy tensor $T_{\mu\nu}$. As discussed
in Ref.~\onlinecite{eeloga,eelogb,eelog1,eelog2}, the entanglement
entropy takes the form $S=\frac{B}{\epsilon}P-a(\theta)\log
P+\mathcal{O}(1)$, where $B/\epsilon$ depends on the UV details,
and the universal coefficient $a(\theta)$ is proportional to
$C_{T}$. The function $a(\theta)$ proposed for entanglement
entropy~\cite{eeloga,eelogb} is also proportional to $f(\theta)$
in our result.

\subsection{Example 2: $Z_N$ SPT-trivial transition}

Now let us still assume the system has a $Z_N$ symmetry, but the
system undergoes a transition between a $2d$ $Z_N$ symmetry
protected topological (SPT) state and a trivial state. Both states
are disordered states of the $Z_N$ symmetry, hence in both states
the ODO $\tO_\cC$ should obey a perimeter law. Our main interest
focuses on the trivial-SPT phase transition, especially the
universal features of $\tO_\cC$ at this transition. This example,
and the next few examples will be described by a class of similar
theories: \beqn \mathcal{S} &=& \int d^2x d\tau \ \sum_{\alpha =
1}^{N_f} \bar{\psi}_\alpha \gamma \cdot (\partial - i N a)
\psi_\alpha \cr\cr &+& m \bar{\psi}\psi + \frac{\ii k}{4\pi} a d a
+ \cdots \label{qedmaster} \eeqn with integer $N_f$ and $N$, and
in general these theories will be labelled as QED$_{(N_f, N, k)}$.
The trivial-SPT transition corresponds to QED$_{(2,1, 0)}$, i.e.
$N_f = 2$, $N = 1$ and $k = 0$~\cite{lulee,Tarun_PRB2013}, plus
Chern-Simons terms of background gauge fields which are not
written explicitly in Eq.~\ref{qedmaster}. The trivial-SPT
transition needs certain fine-tuning to reach the critical point
described by this field theory, hence this field theory is a
multi-critical point between the two states. This multi-critical
point is self-dual~\cite{xudual,mross,seiberg2} and also dual to
the easy-plane deconfined quantum critical
point~\cite{potterdual,xudual1,mengdual,xudual2}. The Dirac
fermion mass term $m$ in Eq.~\ref{qedmaster} is the tuning
parameter between the trivial and SPT phases.

In the theory QED$_{(2,1,0)}$, the current of the U(1) symmetry in
which the microscopic $Z_N$ symmetry is embedded, is $J =
\frac{\ii}{2\pi} \ast da$, and the ODO of the system is given by
Eq.~\ref{znodo}. The angle dependence of the ODO is still give by
Eq.~\ref{angle function}, with $\sigma$ replaced by the
counterpart at the trivial-SPT (multi-)critical point
QED$_{(2,1,0)}$. The universal conductivity can be computed using
various methods such as $1/N_f$ expansion.

\section{Systems with explicit $Z_N^{(1)}$ symmetry}

\subsection{Topological transition at the boundary of a $3d$ SPT
with $Z_N^{(1)} \times \U(1)^{(0)}$ symmetry}

Here we consider an example with an explicit $Z_N^{(1)}$ 1-form
symmetry. The infrared of this example is described by
QED$_{(1,2N,0)}$ of Eq.~\ref{qedmaster}, i.e. it is a single
massless Dirac fermion $\psi$ with charge$-2N$ coupled with a
$\U(1)$ gauge field. In our construction of theory
QED$_{(1,2N,0)}$ we also need a charge$-N$ fermion $\psi'$ in the
background, hence the system only has a $Z_N^{(1)}$ 1-form
symmetry, i.e. the electric flux of the gauge field through any
closed surface is conserved mod $Z_N$. We also demand that the
magnetic flux of the QED$_{(1,2N,0)}$ is conserved, which
corresponds to another $\U(1)^{(0)}$ symmetry. There is a mixed
anomaly between the $Z_N^{(1)}$ and $\U(1)^{(0)}$ symmetries.
Hence the field theory QED$_{(1,2N,0)}$ can be realized at the
boundary of a $3d$ SPT state with $Z_N^{(1)}$ and $\U(1)^{(0)}$
symmetry~\cite{1formphysics}. In the following paragraphs we spell
out this construction of the $3d$ bulk SPT state.~\footnote{This
is one possible construction of the $3d$ bulk, the field theory
QED$_{(1,2N,0)}$ maybe realized as the boundary theory of other
$3d$ 1-form SPT states too.}

To construct the boundary theory QED$_{(1,2N,0)}$, we first
consider a $3d$ bulk with an ordinary photon phase of gauge field
$a_\mu$, and only charge$-N$ and charge$-2N$ fermionic matter
field is dynamical, although all the integer-charge Wilson loops
are allowed in the theory. Hence the system has a $Z_N^{(1)}$
1-form symmetry. All the fermionic matters are in a topologically
trivial band structure in $3d$. Then we bind the Dirac monopole of
$\vec{a}$ with another gauge neutral boson with global
$\U(1)^{(0)}$ conservation, and condense the bound state. The $3d$
bulk is a SPT state with $Z_N^{(1)} \times \U(1)^{(0)}$
symmetry~\cite{1formphysics}. The natural $2d$ boundary of the
system is a $(2+1)d$ photon phase. To create a gauge flux at the
$2d$ boundary, one needs to move a Dirac monopole from outside of
the system, into the $3d$ bulk; since in the $3d$ bulk the bound
state between the Dirac monopole and the $\U(1)^{(0)}$ boson is
condensed, the $2\pi$ magnetic flux at the boundary must also
carry the $\U(1)^{(0)}$ boson. Hence the photons at the $2d$
boundary is the dual of the Goldstone modes of the $\U(1)^{(0)}$
symmetry. Notice that the bulk is fully gapped and has no
spontaneous breaking of the $\U(1)^{(0)}$ symmetry, because the
condensed bound state in the bulk is coupled to the dual gauge
field while carrying the $\U(1)^{(0)}$ charge. The condensate is
still gapped due to the Higgs mechanism.

At the $2d$ boundary, the charge$-2N$ fermion $\psi$ is tuned
close to the transition between a trivial insulator and a Chern
insulator with Chern number $+1$. Due to the fermi-doubling in
$2d$, there must be another massive Dirac cone of $\psi$ in the
band structure that affects the dynamics of $a_\mu$. Hence we need
to design a background band structure of the charge$-N$ fermion
$\psi'$ with Chern number $-2$. The Chern-Simons term of $a_\mu$
generated from $\psi'$ will cancel the Chern-Simons term generated
by the band structure of fermion $\psi$.

Now we have arrived at the theory QED$_{(1,2N,0)}$. The
QED$_{(1,2N,0)}$ is a transition between two different topological
states tuned by the mass of the Dirac fermion $\psi$, these two
topological orders are described by the CS term for $a_\mu$ with
level $k = \pm 2N^2$, which is free of $Z_N^{(1)}$ 1-form symmetry
anomaly. The ODO for the $Z_N^{(1)}$ symmetry is the charge-1
Wilson loop $O_\cC = \exp(\ii \int d\vec{l} \cdot \vec{a})$. In
this case the quantity $\langle( \log O_\mathcal{C})^2 \rangle$ at
the critical point $m = 0$ can be evaluated exactly, based on the
fermion-vortex duality developed
recently~\cite{son2015,maxashvin,wangsenthil1,seiberg1,Cordova_2018}:
\beqn && \mathrm{QED}_{(1,2N,0)} \ \lrar\ \cr\cr &&
\bar{\chi}\gamma\cdot\partial\chi\ \mathrm{coupled\ to\ } Z_{N}\
\mathrm{gauge} \ \mathrm{theory} + \cdots \eeqn The detailed and
exact form of the duality can be found in
Ref.~\onlinecite{Cordova_2018}. The right hand side of the duality
is a Dirac fermion coupled with a $Z_{N}$ gauge field. The duality
relation we will exploit is \beqn J_\chi = \ii \frac{2N}{4\pi}
\ast da, \eeqn where $J_\chi$ is the current carried by $\chi$.
Although $\chi$ is coupled with a $Z_N$ gauge field, since the
$Z_N$ gauge field is gapped, in the infrared the correlation of
$J_\chi$ is identical to that of the free Dirac fermion, and can
be computed exactly: \beqn \left\langle
J_{\chi,\mu}(0)J_{\chi,\nu}(\mathbf{x})\right\rangle
=\frac{1}{8\pi^{2}}\frac{I_{\mu\nu}(\mathbf{x})}{\left|\mathbf{x}\right|^{4}}.
\eeqn One can determine the propagator of the dual gauge field
accordingly. Considering again the $\cC$ in FIG.~\ref{AngleShape},
we find
\begin{flalign}
-\langle(\log
O_{\mathcal{C}})^{2}\rangle=\frac{1}{8N^{2}}\left(\frac{\pi
P}{\epsilon}-f(\theta)\log P\right)+\mathcal{O}(1),
\end{flalign}
where $f(\theta)$ is given in Eq.~\ref{angle function}.

\subsection{QED$_{(N_f, N, k)}$ with explicit $Z_N^{(1)}$ symmetry
and Chern-Simons term}

We consider the theory QED$_{N_f,N,k}$ with large$-N_f$ and level
$k = q N^2$, where $q$ is an integer at the order of $N_f$.
QED$_{(N_f, N, k)}$ with even integer $N_f$, and a CS term with
level $k$ being integer multiple of $N^2$ can be constructed in
$2d$ with $Z_N^{(1)}$ 1-form symmetry~\footnote{We can verify that
the absence of the anomaly associated to the $Z_N$ 1-form symmetry
in this QED theory by considering the its massive phases. For
example, when a positive mass of the Dirac fermion is turned on,
one obtains a U(1) CS theory of level $(q+N_f/2)N^2$. In this
massive phase, the $Z_N$ 1-form symmetry is generated by the anyon
line operator carrying U(1) charge $(q+Nf/2)N$. When $N$ is odd,
we should in fact view the U(1) gauge field $a$ as a spin$_c$
gauge field.  Consequently, this charge-$(q+N_f/2)N$ anyon always
has bosonic self-statistics, which indicates the absence of
anomaly associated with the $Z_N$ 1-form symmetry.  When $N$ is
even, the QED (and its massive phases) intrinsically resides in a
fermionic Hilbert space. The gauge field $a$ is now a regular U(1)
gauge field. In this case, the charge-$(q+N_f/2)N$ anyon can have
either bosonic or fermionic self-statistics depending on the value
of $(q+N_f/2)N$. However, neither case leads to any anomaly
associated to the $Z_N$ 1-form symmetry because the
self-statistics of the charge-$(q+N_f/2)N$ anyon can be made
bosonic by attaching extra neutral fermions in the Hilbert
space.}. At low energy, the dynamics of gauge field is
significantly modified by the one-loop polarization diagram of
fermion $\psi$. In the momentum space, the loop diagram integral
gives \beqn |a_\mu (\vec{p})|^2 \frac{N_f
N^{2}}{16}\frac{\left|p\right|^{2} \delta_{\mu\nu} -
p_{\mu}p_{\nu}}{\left|p\right|} \eeqn which gives an order $N_f$
contribution to the gauge field self-energy. To the leading order
in $1/N_f$, the gauge field propagator in the momentum space is
given by
\begin{flalign}
\frac{16}{N_f N^{2}}\frac{1}{\left|p\right|}
\left(\frac{\cos\hat{\boldsymbol{K}}}{\left|\boldsymbol{K}\right|}
\left(\delta_{\mu\nu}-\zeta\frac{p_{\mu}p_{\nu}}{\left|p\right|^{2}}\right)
+\frac{\sin\hat{\boldsymbol{K}}}{\left|\boldsymbol{K}\right|}
\frac{\varepsilon_{\mu\nu\sigma}p_{\sigma}}{\left|p\right|}\right),
\end{flalign}
where $\left|\boldsymbol{K}\right|,\hat{\boldsymbol{K}}$ denote
the magnitude and the angle of the two-dimensional vector
$\boldsymbol{K}=(1,\frac{-16k}{2\pi N_f N^{2}})$. The Fourier
transformation to real space gives
\begin{flalign}
&\left\langle a_{\mu}(0)a_{\nu}(\mathbf{x})\right\rangle =
\frac{8}{N_f N^{2}}\frac{1}{\pi^{2}\left|\mathbf{x}\right|^{2}}\\
&\times\left(\frac{\cos\hat{\boldsymbol{K}}}{\left|\boldsymbol{K}\right|}
\frac{\delta_{\mu\nu}-\zeta
I_{\mu\nu}(\mathbf{x})}{\left|\mathbf{x}\right|^{2}}+
\frac{\sin\hat{\boldsymbol{K}}}{\left|\boldsymbol{K}\right|}
\frac{\ii\pi}{2}\frac{\varepsilon_{\mu\nu\sigma}x_{\sigma}}
{\left|\mathbf{x}\right|}\right), \notag
\end{flalign}
which has an imaginary part due to the Chern-Simons term. The
parameter $\zeta$ is introduced by gauge fixing.

The ODO for the $Z_N^{(1)}$ symmetry is still the charge-1 Wilson
loop $O_\cC = \exp(\ii \int d\vec{l} \cdot \vec{a})$. As for the
shape of $\cC$ with a sharp corner in FIG.~\ref{AngleShape}, our
calculation leads to the gauge invariant result
\begin{flalign}
-\langle(\log O_{\mathcal{C}})^{2}\rangle=
\frac{8N^{2}N_f}{64k^{2}+\pi^{2}N^{4}N_f^{2}}\left(\frac{\pi
P}{\epsilon}-f(\theta)\log P\right)+\mathcal{O}(1),
\end{flalign}
where $f(\theta)$ is given in Eq.~\ref{angle function}. The
imaginary antisymmetric part of $\left\langle
a_{\mu}a_{\nu}\right\rangle$ does not contribute, and the final
result has the similar form as before. In the large$-N_f$ limit
the universal conductivity of the current $J = \frac{1}{2\pi} \ast
da$ can be computed exactly.

\section{The ``Strange Correlator" of ODO}

Following the argument from Ref.~\onlinecite{xusenthil}, if a
state $|\Omega \rangle$ is the ground state described by a
Lagrangian $\mathcal{L}(\Phi(\bx))$, the matrix elements between
$|\Omega \rangle$ and two different field configurations
$|\Phi(\bx)\rangle$ and $|\Phi'(\bx)\rangle$ is given by the path
integral: \beqn && \langle \Phi(\bx)|\Omega \rangle \langle \Omega
| \Phi'(\bx) \rangle \sim \int_{\Phi(\bx,\tau = -\infty) =
\Phi'(\bx)}^{\Phi(\bx,\tau = +\infty) = \Phi(\bx)} D
\Phi(\bx,\tau) \cr\cr &\times& \exp\left(- \int_{-\infty }^{+
\infty} d\tau d^dx \ \mathcal{L}(\Phi(\bx,\tau)) \right), \eeqn
knowing the matrix element, Ref.~\onlinecite{xusenthil} was able
to derive the ground state wave function based on the Lagrangian
description of various SPT states.

Based on the information of the ground state wave function of SPT
state derived from its Lagrangian, the quantity ``strange
correlator" was introduced and designed to diagnose a SPT state
based on its bulk wave function~\cite{xusc}. Let us assume that
$|0\rangle$ and $|1\rangle$ are the trivial state and SPT state
defined within the same bosonic Hilbert space in a two dimensional
real space, and both systems have the same symmetry. The strange
correlator is the quantity $S(\bx, \bx') = \langle
0|\Phi(\bx)\Phi(\bx')|1\rangle/ \langle 0|1\rangle $, where
$\Phi(\bx)$ is the order parameter of the symmetry that defines
the systems.

For a class of Langrangians $\mathcal{L}$, using the derived wave
functions for both the SPT state $|1\rangle$ and trivial state
$|0\rangle$, one would see that the strange correlator $S(\bx,
\bx')$ cannot have a trivial short range correlation at least for
$d = 2$. Another picture to see this is that, if the Lagrangian
$\mathcal{L}$ has an emergent Lorentz invariant description, after
the space-time rotation, the strange correlator which was purely
defined in space, becomes a space-time correlation function at the
one dimensional spatial interface between $|0\rangle$ and
$|1\rangle$. This picture is similar to the construction of
fractional quantum Hall wave function using conformal
blocks~\cite{cftblock}. Because the spatial interface between
$|0\rangle$ and $|1\rangle$ cannot be trivially gapped, the
strange correlator $S(\bx, \bx')$ must be either long ranged, or
have a power-law. Hence the strange correlator can be viewed as a
tool to diagnose a SPT state based on its bulk wave function, and
it has been shown to be effective for many
examples~\cite{sengupta1,sengupta2,sengupta3,zohar1,zohar2,zohar3,mengstrange,beach,frank}.

ODO is the generalization of correlation functions of 0-form
symmetries. Here we generalize the strange correlator to the ODO
of 1-form symmetry i.e. we evaluate the following quantity \beqn
S(\cC) = \langle 0 | O_\cC | 1 \rangle / \langle 0 | 1 \rangle,
\eeqn where $|0\rangle$ and $|1\rangle$ are trivial state and SPT
state with 1-form symmetry respectively. SPT states protected by
1-form symmetries have attracted great interests in the last few
years~\cite{mcgreevy,formsym6,thorngren2015higher,zhu2019,Cordova2019,xubsre,ye2014,juven,juven2,wen1formspt1,wen1formspt2,jian1form1,1formphysics},
we expect this general question of evaluating strange correlator
of ODO to be a new direction that is worth a deep exploration. In
the current work we consider a typical $3d$ SPT state protected by
the $Z_N^{(1)}$ 1-form symmetry as an example. This SPT state can
be described by the following Lagrangian~\cite{Hsin_2019} \beqn
\mathcal{L} =
 \frac{1}{g} \mathrm{tr}[F_{\mu\nu} F_{\mu\nu}]
+ \frac{\ii \Theta}{8\pi^2} \mathrm{tr}[F \wedge F]. \eeqn $F$ is
the curvature tensor of the $\SU(N)$ gauge field. To guarantee
there is a $Z_N^{(1)}$ 1-form symmetry, we only allow dynamical
(but massive) matter fields of the SU$(N)$ gauge field which
carries an adjoint representation of the gauge field, while closed
Wilson loops with other representations of the gauge field are
still allowed. The SPT state corresponds to $\Theta = 2\pi$, while
the trivial state corresponds to $\Theta = 0$ in the Lagrangian.
The interface between $\Theta = 0$ and $\Theta = 2\pi$ is a $2d$
topological order described by $\SU(N)_1$ Chern-Simons theory with
topological degeneracy. For both $\Theta = 0$ or $2\pi$, the
coupling constant $g$ in the Lagrangian is expected to flow to
infinity under renormalization group, hence the $\Theta-$term is
what remains in the infrared limit. The $\Theta-$term is a total
derivative, hence \beqn && \langle A(\bx)|1 \rangle \langle 1 |
A'(\bx) \rangle \sim \int_{A(\bx,\tau = -\infty) =
A'(\bx)}^{A(\bx,\tau = +\infty) = A(\bx)} D A(\bx,\tau) \cr\cr
&\times& \exp\left(- \int_{-\infty }^{+ \infty} d\tau d^3x \
\mathcal{L}(A)_{g \rightarrow +\infty} \right) \cr\cr &\sim&
\exp\left( \int d^3x \frac{\ii}{4\pi} \mathrm{CS}[A] -
\frac{\ii}{4\pi} \mathrm{CS}[A'] \right), \eeqn Hence the wave
function of the SPT state $|1\rangle$, and the trivial state
$|0\rangle$ (corresponds to $\Theta = 0$) in the limit $g
\rightarrow + \infty$ are schematically \beqn |0\rangle &\sim&
\int DA | A \rangle, \cr \cr |1\rangle &\sim& \int DA \exp \left(
\int d^3x \frac{\ii}{4\pi} \mathrm{CS}[A]\right) | A \rangle.
\eeqn

Now the evaluation of the strange correlator of ODO, which is a
purely $3d$ spatial quantity, is mathematically equivalent to
evaluating world lines of anyons in $(2+1)d$ $\SU(N)_1$ CS field
theory: \beqn S(\cC) \sim \int DA \ \mathrm{tr} [e^{\ii \int_\cC
d\vec{l} \cdot \vec{A}}] \exp\left( \int d^3x \frac{\ii}{4\pi}
\mathrm{CS}[A]\right). \eeqn Then if the ODO is a Wilson loop with
the fundamental representation of the gauge group, and $\cC$
contains two loops with a link, then this evaluation is identical
to the braiding process of two anyons of the $\SU(N)_1$
topological order, and it yields phase $\exp(\ii 2\pi/N^2)$ for
$S(\cC)$.

\section{Discussion}

In this work we studied the behavior of the ``order diagnosis
operator" of 1-form symmetries (for either explicit 1-form
symmetry, or inexplicit 1-form symmetry as a dual of a 0-form
symmetry) at various $(2+1)d$ quantum phase transitions. We
demonstrate that for a class of transitions there is a universal
logarithmic contribution to the ODO arising from the corners of
the loop upon which the ODO is defined. For this class of
transitions, the universal logarithmic contribution is related to
the universal conductivity at the critical points, and in some
cases can be computed exactly using the duality between conformal
field theories.

This logarithmic contribution is similar to the corner
contribution to the entanglement entropy, in fact this relation
can be made exact for free boson/fermion systems~\cite{chengcat}.
For general systems, the ODO associated with certain 1-form
symmetry and the entanglement entropy can be studied in a unified
framework. To study the Renyi entropy, one needs to use the
replica trick, and duplicate $n-$copies of the system. Then the
system is granted an extra ``swapping symmetry" between replica
indices. The Renyi entropy reduces to evaluating the ODO of the
1-form dual of the swapping
symmetry~\cite{Casini_2005,Casini_2009}. Hence we can start with
the duplicated system, and just study the ODO of all the
symmetries of the duplicated system, to extract the information of
both the intrinsic symmetries, and the entanglement entropy
simultaneously. One remark worth making is that, when computing
Renyi entropy for ordinary systems with a Hamiltonian and
translation invariance, there is no interaction between different
duplicated systems, hence each duplicated copy has its own
conservation laws.

In this work we also computed the strange correlator of the 1-form
ODO for a particular example. SPT states protected by 1-form
symmetries have attracted great efforts and interests in the last
few years, and we believe the strange correlator of the 1-form ODO
can be applied to many related systems. We will leave the more
general discussion of this topic to future studies.

The authors thank Wenjie Ji and Yi-Zhuang You for very helpful
discussions. This work is supported by NSF Grant No. DMR-1920434,
the David and Lucile Packard Foundation, and the Simons
Foundation.

{\it Note:} We would like to draw the readers attention to a
closely related work by Yan-Cheng Wang, Meng Cheng and Zi Yang
Meng~\cite{chengdisorder} to appear in the same arXiv listing.

\bibliography{log}

\end{document}